\documentclass[12pt]{article}
\usepackage{amsmath,amsfonts,amssymb}

\usepackage{indentfirst}
\makeatletter
\renewcommand{\@begintheorem}[2]{\begin{trivlist}
\item[\hspace{\labelsep}{\bfseries#1\ #2.}]\itshape}
\renewcommand{\@opargbegintheorem}[3]{\begin{trivlist}
\item[\hspace{\labelsep}{\bfseries#1\ #2\ (#3).}]\itshape}
\renewcommand{\@endtheorem}{\end{trivlist}}
\makeatother
\author{Sergey V.\,Smirnov\thanks{Department of Mathematics and Mechanics, Moscow State University. E-mail: {\tt ssmirnov@higeom.math.msu.su}}}
\title{Semidiscrete Toda lattices}

\def\pa{\partial}
\def\eps{\varepsilon}
\def\phi{\varphi}

\newcommand{\const}{\mathop{\rm const}\nolimits}

\newtheorem{proposition}{\sc Proposition}
\newtheorem{theorem}{\sc Theorem}
\newtheorem{remark}{\sc Remark}

\begin{document}

\maketitle
\begin{abstract}
Integrable cut-off constraints for semidiscrete Toda lattice are studied in this paper. Lax presentation for semidiscrete
analog of the $C$-series Toda lattice is obtained. Nonlocal variables that allow to express symmetries of the infinite semidiscrete lattice are introduced and
cut-off constraints or a certain type compatible with symmetries of the infinite lattice are classified.
\end{abstract}

\section{Introduction}

Various Toda chains and lattices are being examined already for a considerable period of time. One-dimensional Toda chain that describes a system of particles
on a line with exponential interaction between each pair of neighboring particles was introduced
by M.\,Toda in 1967 in his article~\cite{To}. After that Bogoyavlensky~\cite{Bo} introduced generalized one-dimensional Toda chains corresponding to simple Lie
algebras in 1976. In the end of 1970-ties and in the beginning of 1980-ties a series of papers (see~\cite{Mi}--\cite{LSSh}) where
generalized two-dimensional Toda lattices were considered appeared almost at the same time.\footnote{The reference given here is far from being complete, but
the format of this article does not allow us to give full review on the subject.} In 1991 Suris~\cite{Su} studied discrete generalized one-dimensional Toda
chains. This paper is focused on semidiscrete two-dimensional Toda lattices.

It is well-known (e.g. see~\cite{Sh95a} for the details) that in the continuous case the Laplace invariants $h(j)=c(j)-a(j)b(j)-b_y(j)$ of a sequence of hyperbolic
second order differential operators
\begin{equation}
\label{jjj}
{\mathcal L}_j=\pa_x\pa_y+a(j)\pa_x+b(j)\pa_y+c(j),
\end{equation}
linked by Darboux--Laplace transformations
\begin{equation}
\label{kkk}
{\mathcal L}_{j+1}{\mathcal D}_j={\mathcal D}_{j+1}{\mathcal L}_j,
\end{equation}
where ${\mathcal D}_j=\pa_x+b(j)$, satisfy the {\it two-dimensional Toda lattice}:
\begin{equation}
\label{htoda}
(\ln h(j))_{xy}=h(j-1)-2h(j)+h(j+1).
\end{equation}
Using the substitution $h(j)=\exp(q(j+1)-q(j))$ one can rewrite the lattice~(\ref{htoda}) in the form
\begin{equation}
\label{qtoda}
q_{xy}(j)=\exp(q(j+1)-q(j))-\exp(q(j)-q(j-1)),
\end{equation}
and using the substitution $h(j)=u_{xy}(j)$ one obtains the following lattice:
\begin{equation}
\label{utoda}
u_{xy}(j)=\exp(u(j-1)-2u(j)+u(j+1)).
\end{equation}
The latter system (or, more exactly, its reduction defined by the constraints $u(-1)=u(r)=-\infty$ for some natural $r$) is a particular case of the so-called
{\it systems of exponential type}, i.e. the equations of the form ${\mathbf u}_{xy}=\exp(K\mathbf u)$, where $K$ is a constant matrix, and $\exp(K\mathbf u)$
denotes the vector such that its $j$-th coordinate is equal to the exponent of the $j$-th coordinate of the vector $K{\mathbf u}$. A system of exponential
type is proven to be Darboux integrable (i.e. a full set of $y$-integrals of this system is obtained) if and only if matrix $K$
is the Cartan matrix of a simple Lie algebra (see papers~\cite{ShJa,LSSh}). Moreover, systems of exponential type corresponding to simple Lie algebras were
explicitly integrated in~\cite{L}. Such complete description of integrable systems of exponential type is based on the notion of {\it characteristic algebra}
of a system of equations. It is proved that the existence of a full set of $y$-integrals of a system is equivalent to finite dimensionality of its
characteristic algebra; the latter condition holds for a system of exponential type if and only if matrix $K$ is the Cartan matrix of a Lie algebra of
series $A-D$ or the Cartan matrix of an exceptional Lie algebra. These systems are called {\it generalized} or {\it finite} two-dimensional Toda lattices.
Lax presentations for all generalized two-dimensional Toda lattices were obtained in~\cite{MOP} using methods of theory of Lie algebras. In the most simple
case of the boundary condition $u(-1)=u(1)=-\infty$ the system~(\ref{utoda}) is reduced to the famous Liouville equation. This hyperbolic PDE had been explicitly
integrated by Liouville~\cite{Liouv} in 1853.

Despite the fact that complete description of all explicitly integrable finite two-dinensional Toda lattices had been provided, some problems still remained
unsolved. In particular, it was not clear whether $D$-series two-dimensional Toda lattice is a reduction of an $A$-series lattice until this
problem was affirmatively resolved by Habibullin~\cite{Ha05} in 2005 ($B$-series and $C$-series lattices are reductions of an $A$-series Toda lattice defined
by involute constraints $h(-j)=h(j-1)$ and $h(-j)=h(j)$ respectively). Besides this, explicit integrability and existence of a full set of $y$-integrals are not
the only attributes of a concept of integrable equation in our days. A new idea based on the notion of integrability as existence of a Lax presentation was
proposed in~\cite{Ha05}. Although this approach to generalized Toda lattices is systematic since it provides an integrable cut-off constraint {\it together}
with a Lax pair (it was an additional problem to find a Lax pair for a finite Toda lattice before), it appears to be not very effective.

One more approach to the problem of classification of integrable cut-off constraints for two-dimensional Toda lattice developed in~\cite{GH} is based on the
study of boundary conditions compatible with symmetries of the infinite Toda lattice.

The structure of this article is as follows. Habibullin's direct approach that allows to find integrable cut-off constraints together with corresponding
Lax presentation is described in the Section~\ref{contlax}. Section~\ref{contsymm} contains Shabat's method~\cite{Sh95b} of obtaining symmetries for
two-dimensional infinite Toda and the symmetry approach to the problem of classification of integrable cut-off constraints for two-dimensional Toda
lattice developed by G\"urel and Habibullin~\cite{GH} in the continuous case. Semidiscrete Toda lattice is introduced in the Section~\ref{semidisc} (here
we follow the article~\cite{AS} by Adler and Startsev). Lax presentation for the semidiscrete analog of a $C$-series Toda lattice is obtained in the
Section~\ref{disclax}.\footnote{When this article was almost completed the author discovered the preprint on an article~\cite{HZY} where, in particular,
Lax presentation for semidiscrete $C$-series Toda lattice is obtained.} Symmetry approach to the problem of classification of integrable cut-off constraints
for the Toda lattice in the semidiscrete case is developed in the Section~\ref{discsymm}.

\section{Lax pair approach in the continuous case}\label{contlax}

Before going over to Habibullin's straightforward method of obtaining Lax presentations for the finite two-dimensional Toda lattices we'll display
explicitly the cut-off constraints corresponding to $A-D$-series Toda lattices in terms of the variables $q(j)$. $A$-series Toda lattice is
defined by the trivial boundary conditions
\begin{equation}
\label{aseries}
q(-1)=\infty, \quad q(r+1)=-\infty
\end{equation}
for some $r\in\mathbb N$. $B$-series lattice is defined the following cut-off constraint:
\begin{equation}
\label{bseries}
q(0)=0, \quad q(r+1)=-\infty
\end{equation}
for some $r\in\mathbb N$. $C$-series lattices are generated by cut-off constraints of the form
\begin{equation}
\label{cseries}
q(0)=q(1), \quad q(r+1)=-\infty,
\end{equation}
and $D$-series lattices are generated by a more complicate cut-off constraint:
\begin{equation}
\label{dseries}
q(0)=-\ln\left(e^{q(2)}-\frac{q_x(1)q_y(1)}{2\sinh q(1)}\right), \quad q(r+1)=-\infty.
\end{equation}

It is well-known that Darboux-Laplace transformations provide a Lax presentations for the infinite two-dimensional Toda lattice:
equations~(\ref{qtoda}) are equivalent to the compatibility conditions of the following linear system of equations:
\begin{eqnarray}
\label{lax1}
\left\lbrace
\begin{array}{l}
\psi_{x} (j)=\psi (j+1)+q_x(j)\psi (j)\\
\psi_{y} (j)=-h(j-1)\psi (j-1)
\end{array}
\right.,
\end{eqnarray}
where $h(j)=\exp(q(j+1)-q(j))$. We are interested in finite Toda lattices. Therefore we need to obtain Lax presentations for Toda equations reduced
by the above cut-off constraints. Consider an arbitrary cut-off constraint of the form
\begin{equation}
\label{ggg}
q(-1)=F(q(0),q(1),\dots,q(k))
\end{equation}
for some $k\in\mathbb N$. Since the equation for $\psi_y (j)$ contains $psi(-1)$, it is rather natural to make an attempt to express $\psi(-1)$ via
the dynamic variables $\psi(0),\psi(1),\dots,\psi(k)$ in a way
consistent with the constraint~(\ref{ggg}). This would have allowed one to obtain a Lax presentation for the reduced lattice. However, such
procedure works only for the trivial boundary conditions and this approach produces a Lax pair only for an $A$-series Toda lattice (see~\cite{Ha05}).
This obstacle led Habibullin to the other idea: since the infinite Toda lattice~(\ref{qtoda}) is symmetric about the interchange
of variables $x\leftrightarrow y$, it also admits another Lax presentation:
\begin{eqnarray}
\label{lax2}
\left\lbrace
\begin{array}{l}
\phi_{x}(j)=-h(j-1)\phi (j-1)\\
\phi_y(j)=\phi (j+1)+q_y(j)\phi (j)
\end{array}
\right..
\end{eqnarray}
This means that one may seek for a closure of a Lax pair~(\ref{lax1}) using not only this Lax pair but to find a common closure for both Lax pairs,
that is, to eliminate the variables $\psi(-1)$ and $\phi(-1)$ using the {\it whole} set of variables
$$
\psi(0),\psi(1),\dots,\psi(r),\phi(0),\phi(1),\dots,\phi(r)
$$
in a way consistent with the constraint~(\ref{ggg}). This approach provides Lax pairs for all generalized Toda lattices of the series $A-D$; moreover,
in some cases these Lax pairs do not coincide with the ones that were obtained using Lie algebraic methods in\cite{MOP}. The following
Proposition (see~\cite{Ha05}) holds.
\begin{proposition}
Let $k<k_1$ and $m<m_1$ be integers. Consider vectors
$$
\Psi=\left(\psi(k+1),\psi(k+2),\dots,\psi(k_1-1)\right)^t,\quad\Phi=\left(\phi(m+1),\phi(m+2),\dots,\phi(m_1-1)\right)^t;
$$
then all generalized two-dimensional Toda lattices of the series $A-D$ admit Lax presentation of the following form:
\begin{eqnarray}
\label{hhh}
\left(
\begin{array}{c}
\Psi\\
\Phi
\end{array}
\right)_x=\left(
\begin{array}{c|c}
A & K\\
\hline
M & C
\end{array}
\right)\left(
\begin{array}{c}
\Psi\\
\Phi
\end{array}
\right),\quad
\left(
\begin{array}{c}
\Psi\\
\Phi
\end{array}
\right)_y=\left(
\begin{array}{c|c}
B & L\\
\hline
N & D
\end{array}
\right)\left(
\begin{array}{c}
\Psi\\
\Phi
\end{array}
\right),
\end{eqnarray}
where matrices $A,B,\dots, N$ look as follows:
\begin{eqnarray}
\nonumber
A=\left(
\begin{array}{ccccc}
q_x (k+1) & 1 & 0 & \dots & 0\\
0 & q_x (k+2) & 1 & \dots & 0\\
\vdots & & \ddots & \ddots & \\
0 & 0 & & \dots & 1\\
A_{k+1} & A_{k+2} & A_{k+3} & \dots & A_{k_1-1}
\end{array}
\right),\quad
K=\left(
\begin{array}{cccc}
0 & 0 & \dots & 0\\
0 & 0 & \dots & 0\\
\vdots & & \ddots &\\
0 & 0 & \dots & 0\\
K_{m+1} & K_{m+2} &\dots & K_{m_1-1}
\end{array}
\right),
\end{eqnarray}
\begin{eqnarray}
\nonumber
B=\left(
\begin{array}{ccccc}
B_{k+1} & B_{k+2} & \dots & 0 & B_{k_1-1}\\
-h(k+1) & 0 & \dots & 0 & 0\\
\vdots & \ddots & \ddots & \\
0 & 0 & \dots & 0 & 0\\
0 & 0 & \dots & -h(k_1-2) &0
\end{array}
\right),\quad
L=\left(
\begin{array}{cccc}
L_{m+1} & L_{m+2} & \dots & L_{m_1-1}\\
0 & 0 & \dots & 0\\
\vdots & & \ddots &\\
0 & 0 & \dots & 0\\
0 & 0 & \dots & 0
\end{array}
\right),
\end{eqnarray}
\begin{eqnarray}
\nonumber
C=\left(
\begin{array}{ccccc}
C_{m+1} & C_{m+2} & \dots & 0 & C_{m_1-1}\\
-h(m+1) & 0 & \dots & 0 & 0\\
\vdots & \ddots & \ddots & \\
0 & 0 & \dots & 0 & 0\\
0 & 0 & \dots & -h(m_1-2) &0
\end{array}
\right),\quad
M=\left(
\begin{array}{cccc}
M_{k+1} & M_{k+2} & \dots & M_{k_1-1}\\
0 & 0 & \dots & 0\\
\vdots & & \ddots &\\
0 & 0 & \dots & 0\\
0 & 0 & \dots & 0
\end{array}
\right),
\end{eqnarray}
\begin{eqnarray}
\nonumber
D=\left(
\begin{array}{ccccc}
q_y (m+1) & 1 & 0 & \dots & 0\\
0 & q_y (m+2) & 1 & \dots & 0\\
\vdots & & \ddots & \ddots & \\
0 & 0 & & \dots & 1\\
D_{m+1} & D_{m+2} & D_{m+3} & \dots & D_{m_1-1}
\end{array}
\right),\quad
N=\left(
\begin{array}{cccc}
0 & 0 & \dots & 0\\
0 & 0 & \dots & 0\\
\vdots & & \ddots &\\
0 & 0 & \dots & 0\\
N_{k+1} & N_{k+2} &\dots & N_{k_1-1}
\end{array}
\right).
\end{eqnarray}
\end{proposition}
\begin{remark}
\rm
One of the advantages of this approach is that cut-off constraints on the left edge and on the right edge are independent from each other.
Therefore not only generalized Toda lattices of the series $A-D$ admit Lax presentation of the form~(\ref{hhh}), but any finite Toda lattice
defined by any of the constraints~(\ref{aseries})--(\ref{dseries}) on the left and on the right edges also admits such Lax presentation.
Moreover, there are examples of boundary conditions~(\ref{ggg}) such that corresponding system is not a system of exponential
type although it admits Lax presentation of the form~(\ref{hhh}) (see~\cite{Ha05}).
\end{remark}
\begin{remark}
\rm
Probably the lattice~(\ref{qtoda}) admits infinitely many cut-off constraints that are integrable in this sense. However, complete
classification of all such cut-off constraints seems to be absolutely impossible due to the fact that one has to solve a huge system of
differential-difference equations in order to determine the free entries of the matrices $A,B,\dots,N$. Besides this, Habibullin's approach is
not very effective in the sense that even if one is interested in a Lax pair for a certain cut-off constraint, it is quite unclear how to
choose the integers $k$, $m$, $k_1$ and $m_1$ and there is absolutely no indication on how many entries in each of the above matrices
should be left undetermined.
\end{remark}

Functions $\psi(j)$ and $\phi(j)$, satisfying the equations~(\ref{lax1}) and~(\ref{lax2}), are also solutions of the hyperbolic equations
${\mathcal L}^{\psi} (j)\psi(j)=0$ and ${\mathcal L}^{\phi} (j)\phi(j)=0$ respectively, where the operators ${\mathcal L}^{\psi}(j)$
and ${\mathcal L}^{\phi}(j)$ are defined by the equations:
$$
{\mathcal L}^{\psi}(j)=\pa_x\pa_y-q_x(j)\pa_y+h(j-1),\quad {\mathcal L}^{\phi}(j)=\pa_x\pa_y-q_y(j)\pa_x+h(j-1).
$$
Laplace invariants $h(j)=c(j)-a(j)b(j)-b_y(j)$ and $k(j)=c(j)-a(j)b(j)-a_x(j)$ of these operators are as follows:
\begin{equation}
\label{lll}
h^{\psi}(j)=h(j),\quad k^{\psi}(j)=h(j-1),\quad h^{\phi}(j)=h(j-1),\quad k^{\phi}(j)=h(j).
\end{equation}
It is easy to verify that in terms of the variable $h$ the cut-off constraints corresponding to the series $B$ and $C$ are nothing but involutions $h(-j)=h(j+1)$
and $h(-j)=h(j)$ respectively. Combining this with~(\ref{lll}), one obtains the following identities for a $B$-series lattice:
$$
h^{\phi}(0)=h(-1)=h(2)=h^{\psi}(2),\quad k^{\phi}(0)=h(0)=h(1)=k^{\psi}(2).
$$
It is well known that two linear hyperbolic operators have the same Laplace invariants if and only if they are linked by a gauge transformation. In terms of
wave functions this means that there exists a multiplier $R=R(x,y)$ such that $\phi(0)=R\psi(2)$ (i.e for any solution $\psi(2)$ of one of this equations the
function $\phi(0)=R\psi(2)$ is the solution to the other equation). Similarly, one establishes that
$$
h^{\psi}(1)=h(1)=h(0)=h^{\phi}(1),\quad k^{\psi}(1)=h(0)=h(1)=k^{\phi}(1);
$$
this means that there exists another multiplier $S=S(x,y)$ such that $\psi(1)=S\phi(1)$. The identities obtained are nothing but a particular case of
Habibullin's general idea to consider common reductions of two Lax pairs for the infinite Toda lattice at the left edge. In the case of a $C$-series
Toda lattice one immediately arrives at the following identities:
\begin{align}
\notag
&h^{\phi}(0)=h(-1)=h(1)=h^{\psi}(1),\quad &&k^{\phi}(0)=h(0)=k^{\psi}(1),\\
\notag
&h^{\psi}(0)=h(0)=h^{\phi}(1),\quad &&k^{\psi}(0)=h(-1)=h(1)=k^{\phi}(1).
\end{align}
Hence there exist the multipliers $R$ and $S$ such that $\phi(0)=R\psi(1)$, $\psi(0)=S\phi(1)$. It is also a particular case of the above Habibullin's
approach. The cut-off constraint corresponding to $D$-series lattices is more complicate; in this case the two Lax pairs are not gauge equivalent in any
point. The examples of boundary conditions for the series $B$ and $C$ demonstrate that the idea to consider two Lax pairs together and to look for their
common reduction is rather natural.

\section{Symmetry approach in the continuous case}\label{contsymm}

Another approach to the problem of integrable cut-off constraints for the two-dimensional Toda lattice proposed by G\"urel and Habibullin~\cite{GH} is
based on the study of boundary conditions compatible with symmetries of the infinite lattice. However, symmetries of the infinite two-dimensional Toda
lattice cannot be expressed in terms of the dynamic variables (on the contrary to the one-dimensional case). In the two-dimensional case one has to
introduce nonlocal variables in order to define symmetries of the infinite Toda lattice.

Denote $q_x(j)$ by $b(j)$ (this variable has nothing in common with coefficients of hyperbolic operators from the previous Section). Hence the Toda
lattice can be rewritten as follows:
\begin{equation}
\label{btoda}
b_y(j)=h(j)-h(j-1).
\end{equation}
Define the nonlocal variables $b^{(1)}(j)$ by the identities
\begin{equation}
\label{all}
\pa_x b (j)=b^{(1)}(j)-b^{(1)}(j-1),\quad\pa_y b^{(1)} (j)=\pa_x h(j).
\end{equation}
Compatibility of these equations is provided by the Toda lattice~(\ref{btoda}):
$$
\pa_y\pa_x b (j)=\pa_y \left(b^{(1)}(j)-b^{(1)}(j-1)\right)=\pa_x \left(h(j)-h(j-1)\right)=\pa_x\pa_y b(j).
$$
Further on, it is necessary to determine the derivatives $\pa_x b^{(1)}(j)$, but this, however, requires to introduce nonlocalities $b^{(2)} (j)$
of higher order. Set
$$
\pa_x b^{(1)} (j)=b^{(1)}(j)\left(b(j+1)-b(j)\right)+b^{(2)}(j)-b^{(2)}(j-1);
$$
hence the compatibility condition $\pa_y\pa_x b^{(1)}(j)=\pa_x\pa_y b^{(1)}(j)$ obviously leads to the following identity:
$$
\pa_y b^{(2)}(j)=h(j)b^{(1)}(j+1)-h(j+1)b^{(1)}(j).
$$
The following Proposition is established straightforwardly by induction.
\begin{proposition}
Nonlocal variables $b^{(k)}(j)$, where $k=2,3,\dots$, satisfy the following identities:
\begin{eqnarray}
\label{uuu}
\left\lbrace
\begin{array}{lll}
\pa_y b^{(k)}(j)&=&h(j)b^{(k-1)}(j+1)-h(j+k-1)b^{(k-1)}(j)\\
\pa_x b^{(k)}(j)&=&b^{(k)}(j)\left(b(j+k)-b(j)\right)+b^{(k+1)}(j)-b^{(k+1)}(j-1)
\end{array}
\right.,
\end{eqnarray}
Compatibility of these equations for each $k$ is provided by the first of them, but for $k+1$.
\end{proposition}

The formulas~(\ref{uuu}) were obtained by A.\,B.\,Shabat~\cite{Sh95b} in order to derive the hierarchy of
symmetries of the infinite two-dimensional Toda lattice. The following Proposition holds.
\begin{proposition}
The flows
\begin{eqnarray}
\label{symm2}
q_t(j)&=&b^2 (j)+b^{(1)}(j)+b^{(1)}(j-1),\\
\nonumber
q_t(j)&=&b^3 (j)+b^{(2)}(j)+b^{(2)}(j-1)+b^{(2)}(j-2)+\\
\label{symm3}
&+&b^{(1)}(j)\left(2b(j)+b(j+1)\right)+b^{(1)}(j-1)\left(2b(j)+b(j-1)\right)
\end{eqnarray}
define symmetries of the lattice~(\ref{btoda}).
\end{proposition}

The nonlocal variables $b^{(k)}(j)$ also provide an opportunity to obtain first integrals for the two-dimensional Toda lattice~(\ref{qtoda}).
Indeed, it it easy to verify that the functions
$$
I_2=\sum\limits_{j\in\mathbb Z}\left(b^2 (j)+2b^{(1)}(j)\right)\quad\hbox{è}\quad
I_3=\sum\limits_{j\in\mathbb Z}\left(b^3 (j)+3b^{(1)}(j)(b(j)+b(j+1))+3b^{(2)}(j)\right)
$$
are $y$-integrals for the lattice~(\ref{qtoda}). Symmetries and integrals of higher order are expressed in terms of the
nonlocal variables $b^{(k)}(j)$ of higher grading, but we are not going to discuss them in this article.

The main idea of the article~\cite{GH} is to examine what finite Toda lattices admit symmetries and, in particular,
to classify all cut-off constraints, compatible with the symmetries~(\ref{symm2}),(\ref{symm3}). Straightforward calculations
show that the symmetry~(\ref{symm2}) is compatible only with trivial boundary condition (i.e. only $A$-series Toda lattices
admit symmetry~(\ref{symm2})) and that the symmetry~(\ref{symm3}) is compatible, for example, with the cut-off constraints
corresponding to the series $A$, $B$ and $C$. It appears that all boundary conditions for the Toda lattice that are known
to be integrable in some sense are compatible with the symmetry~(\ref{symm3}). In order to formulate the corresponding
theorem one has to introduce the new set of dynamic variables (as well as the new set of nonlocal variables) such that
the symmetries~(\ref{symm2}) are~(\ref{symm3}) are expressed in a more compact way in their terms.

Denote $u=e^{-q(-1)}$ and $v=e^{q(0)}$; then the symmetries~(\ref{symm2}) and~(\ref{symm3}) can be rewritten respectively as follows:
\begin{align}
\label{usymm2}
&u_t=-u_{xx}-2ru,\quad &&v_t=v_{xx}+2rv,\\
\label{usymm3}
&u_t=u_{xxx}+3ru_x-3su+3r_x u,\quad &&v_t=v_{xxx}+3rv_x+3sv,
\end{align}
where $r=b^{(1)}(0)$ and $s=b^{(2)}(0)+r(\ln v)_x$ are the new nonlocal variables that satisfy the relations
$$
r_y=(uv)_x,\quad s_y=(uv_x)_x.
$$

The following theorem is the main result of the article~\cite{GH}.
\begin{theorem}
If a cut-off constraint of the form $u=F(v,v_x,v_y,v_{xy})$ is compatible with the symmetry~(\ref{usymm2}), then it is trivial: $u=0$.
A cut-off constraint $u=F(v,v_x,v_y,v_{xy})$ compatible with the symmetry~(\ref{usymm3}) has one of the following forms:

i) $u=a$, where $a=\const$;

ii) $u=av$, where $a=\const$;

iii) $u=\frac{v_{xy}}{a-v^2}+\frac{vv_x v_y}{(a-v^2)^2}$, where $a=\const$.
\end{theorem}

These cut-off constraints correspond with the well-known boundary conditions for the two-dimensional Toda lattice, corresponding the the
Lie algebras of the series $A$--$D$. Cut-off constraints of more general form provide all boundary conditions for the
Toda lattice that were considered in literature (see~\cite{GH}).

\section{Infinite lattice in the semidiscrete case}\label{semidisc}

In the semidiscrete case as well as in the continuous case the Laplace invariants of a sequence of hyperbolic second order operators
linked by Darboux-Laplace transformations satisfy the system of differential-difference equations
called {\it the semidiscrete Toda lattice}. We'll follow the article~\cite{AS} and use this idea to obtain the semidiscrete Toda equations.

Consider a sequence of hyperbolic differential-difference operators
$$
{\cal L}_j=\pa_x T+a_n(j)\pa_x+b_n(j) T+c_n(j),
$$
where $a_n(j)$, $b_n(j)$ and $c_n(j)$ are functions depending on discrete variable $n\in\mathbb Z$ and on continuous variable $x\in\mathbb R$
and where $T$ is a shift operator: $T\psi_n (x)=\psi_{n+1}(x)$. Obviously, the operator ${\cal L}_j$ can be factorized in two different ways:
$$
{\cal L}_j=(\pa_x+b_n(j))(T+a_n(j))+a_n(j)k_n(j)=(T+a_n(j))(\pa_x+b_{n-1}(j))+a_n(j)h_n(j),
$$
where $k_n(j)=\frac{c_n(j)}{a_n(j)}-(\ln a_n(j))'_x-b_n(j)$ and
$h_n(j)=\frac{c_n(j)}{a_n(j)}-b_{n-1}(j)$ are {\it the Laplace invariants}
of differential-difference operator ${\cal L}_j$.

Suppose any two neighboring operators ${\cal L}_j$ and ${\cal L}_{j+1}$ are linked by {\it a Darboux-Laplace transformation}, that is, satisfy
the following relation:
$$
{\cal L}_{j+1}{\cal D}_j={\cal D}_{j+1}{\cal L}_j,
$$
where ${\cal D}_j=\pa_x+b_{n-1}(j)$. This operator relation can be rewritten in terms of coefficients as follows:
\begin{eqnarray}
\nonumber
\left\lbrace
\begin{array}{l}
k_n (j+1)=h_n(j)\\
\left(\ln\frac{h_n(j)}{h_{n+1}(j)}\right)'_x=h_{n+1}(j+1)-h_{n+1}(j)-h_n(j)+h_n(j-1)
\end{array}
\right..
\end{eqnarray}
Using the new variables $q_n(j)$ that are introduced by the equations $h_n(j)=\exp(q_{n+1}(j+1)-q_n(j))$, one obtains {\it the semidiscrete Toda lattice}:
\begin{equation}
\label{todaq}
q_{n,x}(j)-q_{n+1,x}(j)=\exp(q_{n+1}(j+1)-q_n (j))-\exp(q_{n+1}(j)-q_n(j-1)).
\end{equation}

Similarly to the continuous case it is natural to examine integrable reductions of the semidiscrete Toda lattice.
Trivial boundary conditions $h_n(-1)=h_n(r)=0$ lead to the system that should be called {\it the semidiscrete lattice
corresponding to the $A$-series Lie algebra}. In terms of the variable $q$ this reduction is defined by the cut-off
constraint $q_n(-1)=\infty$, $q_n (r+1)=-\infty$. Now we'll find out what involutions does the semidiscrete lattice
in terms of the variables $h$, i.e. the system
\begin{equation}
\label{sdhtoda}
\left(\ln\frac{h_n(j)}{h_{n+1}(j)}\right)'_x=h_{n+1}(j+1)-h_{n+1}(j)-h_n(j)+h_n(j-1),
\end{equation}
admit. It is easy to verify that the reflection $h_n(-j)=h_{n+j+c}(j-d)$ defines a reduction of the lattice~(\ref{sdhtoda})
if and only if $d=-2c$. This means that the situation in the semidiscrete case differs from the one in the continuous
case since in the continuous case reflections about both semi-integer and integer points define reductions of the Toda
lattice (these reductions correspond to the series $B$ and $C$ respectively). If $c=-1$, then one obtains the cut-off constraint $h_n(-j)=h_{n+j-1}(j-2)$
(this equation should be satisfied for all $n\in\mathbb Z$). For $j=2$ one arrives at the following cut-off constraint in terms of the variable $q$:
\begin{equation}
\label{todac}
q_n (-1)-q_{n-1}(-2)=q_{n+1}(1)-q_n(0).
\end{equation}
The lattice~(\ref{todaq}) satisfying the cut-off constraint~(\ref{todac}) on the left edge and the trivial boundary condition
$q_n(r+1)=-\infty$ for a certain $r\in\mathbb N$ on the right edge is called {\it semidiscrete Toda lattice corresponding to the $C$-series}. It is
easy to verify that this system converges to the generalized two-dimensional $C$-series Toda lattice in the continuum limit.

Straightforward calculation shows that the boundary condition~(\ref{todac}) defines an $n$-integral of the semidiscrete lattice. Indeed, the function
\begin{equation}
\label{nint}
\mu(x)=q_{n,x}(-1)+q_{n,x}(0)-h_n(0),
\end{equation}
where $q_n(j)$ is a solution of a $C$-series semidiscrete Toda lattice, does not depend on $n$.
\begin{remark}
\rm
The existence of the discrete-time conservation law~(\ref{nint}) is equivalent to the cut-off constraint~(\ref{todac}). Indeed, one can easily verify
that if the value~(\ref{nint}) does not depend on $n$ for a certain solution $q_n (j)$ of the infinite lattice~(\ref{todaq}), then the
relation~(\ref{todac}) holds. However the function $\mu$ may be different for various solutions of the Toda lattice.
\end{remark}

\section{Lax pair for the semidiscrete $C$-series lattice}\label{disclax}

Exactly as in the continuous case, Darboux-Laplace transformations allow to obtain Lax presentations for the semidiscrete Toda lattice~(\ref{todaq}).
It is easy to verify that the equations~(\ref{todaq}) are equivalent to the compatibility conditions for the following linear system of equations:
\begin{eqnarray}
\label{dlax1}
\left\lbrace
\begin{array}{l}
\psi_{n,x} (j)=q_{n,x}(j)\psi_n (j)+\psi_n (j+1)\\
\psi_{n+1} (j)=\psi_n (j)+h_n(j-1)\psi_n(j-1)
\end{array}
\right..
\end{eqnarray}
Darboux-Laplace transformations on discrete variable (i.e. transformations with $D$-operator being discrete) also lead to Lax presentation for the
semidiscrete lattice and this Lax presentation is not equivalent to the first one:
\begin{eqnarray}
\label{dlax2}
\left\lbrace
\begin{array}{l}
\phi_{n,x} (j)=\exp(q_{n+1}(j)-q_{n+1}(j-1))\phi_n (j-1)-\phi_n (j)\\
\phi_{n-1} (j)=-\exp(q_{n}(j)-q_{n+1}(j))\phi_n (j)+\phi_n(j+1)
\end{array}
\right..
\end{eqnarray}
Wave functions $\psi_n (j)$ and $\phi_n(j)$ that satisfy the equations~(\ref{dlax1}) and~(\ref{dlax2}) are solutions of the hyperbolic
differential-difference equations ${\mathcal L}^{\psi} (j)\psi_n(j)=0$ and ${\mathcal L}^{\phi} (j)\phi_n(j)=0$ respectively, where
\begin{align}
\notag
&{\mathcal L}^{\psi}(j)=\pa_x T-\pa_x-q_{n+1,x}(j)T+q_{n,x}(j)-h_{n}(j),\\
\notag
&{\mathcal L}^{\phi}(j)=\pa_x T^{-1}+\exp(q_n(j)-q_{n+1}(j))\pa_x+T^{-1}+\exp(q_n(j)-q_{n+1}(j))-\exp(q_n(j)-q_n(j-1)).
\end{align}
Laplace invariants of these operators are as follows:
\begin{equation}
\label{mmm}
h^{\psi}_n(j)=h_n(j),\quad k^{\psi}_n(j)=h_n(j-1),\quad h^{\phi}_n(j)=h_n(j-1),\quad k^{\phi}_n(j)=h_{n+1}(j).
\end{equation}
One can easily verify that the cut-off constraint $h_n(-j)=h_{n+j-1}(j-2)$ corresponding to the $C$-series together with the
condition~(\ref{mmm}) lead to the following relations:
$$
h_n^{\phi}(-1)=h_n(-2)=h_{n+1}(0)=h_{n+1}^{\psi}(0),\quad k_n^{\phi}(-1)=h_{n+1}(-1)=k_{n+1}^{\psi}(0).
$$
Laplace invariants of two linear hyperbolic differential-difference operators are equal if and only if these operators are gauge-equivalent (exactly
as in the continuous case). In terms of wave functions this means that there exists a function $R_n=R_n(x)$ such that $\phi_n(-1)=R_n\psi_{n+1}(0)$.
Similarly, the following relations hold:
$$
h_n^{\psi}(-1)=h_n(-1)=h_n^{\phi}(0),\quad k_n^{\psi}(-1)=h_n(-2)=h_{n+1}(0)=k_n^{\phi}(0).
$$
Hence there exists a function $S_n=S_n(x)$ such that $\psi_n(-1)=S_n\phi_n(0)$. Therefore, the $C$-series cut-off constraint makes the two Lax
pairs~(\ref{dlax1}),(\ref{dlax2}) gauge-equivalent at one point. According to the general Habibullin's approach it allows to obtain a Lax presentation
for the finite lattice by considering common closure of two Lax pairs for the infinite lattice. More precisely, the following Proposition holds.

\begin{proposition}
Toda lattice~(\ref{todaq}) with the boundary condition~(\ref{todac}) on the left edge and with the trivial boundary condition $q_n(r+1)=-\infty$ for a
certain $r\geqslant 1$ on the right edge is equivalent to the compatibility condition for the following linear system:
\begin{align*}
\pa_x (\Psi)&=A\Psi, & &T(\Psi)=B\Psi+L\Phi,\\
\pa_x (\Phi)&=M\Psi+C\Phi, & &T^{-1}(\Phi)=D\Phi,
\end{align*}
where
$$
\Psi=\begin{pmatrix}
\psi(0)\\
\psi(1)\\
\vdots\\
\psi(r)
\end{pmatrix},\quad\Phi=\begin{pmatrix}
\phi(0)\\
\phi(1)\\
\vdots\\
\phi(r)
\end{pmatrix},
$$
and matrices are as follows:
$$
A=\begin{pmatrix}
p_n & 0 & 0 &\dots & 0 & 0\\
\nu\cdot a_{n+1} (1) & -1 & 0 &\dots & 0 & 0\\
0 & a_{n+1} (2) & -1 &\dots & 0 & 0\\
\vdots && \ddots & \ddots & 0 & 0\\
0 & 0 & 0 & \dots & -1 & 0\\
0 & 0 & 0 & \dots & a_{n+1} (r) & -1
\end{pmatrix},
\quad B=\begin{pmatrix}
b_n(0) & \nu^{-1} & 0 & 0 &\dots & 0\\
0 & b_n(1) & -1 & 0 &\dots & 0\\
0 & 0 & b_n(2) & -1 &\dots & 0\\
\vdots &&&\ddots & \ddots &\\
0 & 0 & 0 & 0 &\dots & -1\\
0 & 0 & 0 & 0 &\dots & b_n(r)
\end{pmatrix},
$$
$$
C=\begin{pmatrix}
q_{n,x}(0) & 1 & 0 & \dots & 0\\
0  & q_{n,x}(1) & 1 &\dots & 0\\
\vdots && \ddots &\ddots&\\
0 & 0 & &\dots & 1\\
0 & 0 & &\dots & q_{n,x}(r)
\end{pmatrix},
\quad D=\begin{pmatrix}
1 & 0 & \dots & 0 & 0\\
h_n(0) & 1 & \dots & 0 & 0\\
\vdots & \ddots & \ddots & 0 & 0\\
0 & 0 & \dots & 1 & 0\\
0 & 0 & \dots & h_n(r-1) & 1
\end{pmatrix},
$$
$$
L=\begin{pmatrix}
f_n & 0 & 0 & \dots & 0\\
0 & 0 & 0 &\dots & 0\\
\vdots &&&\ddots &\\
0 & 0 & 0 & \dots & 0
\end{pmatrix},\quad
M=\begin{pmatrix}
g_n & 0 & 0 & \dots & 0\\
0 & 0 & 0 & \dots & 0\\
\vdots &&&\ddots &\\
0 & 0 & 0 &\dots & 0
\end{pmatrix}.
$$
Matrix elements are defined as follows:
\begin{align*}
&a_n (j)=\exp(q_n(j)-q_n(j-1)), & &b_n (j)=-\exp(q_n(j)-q_{n+1}(j)),\\
&f_n=(-1)^n\exp(-q_n(-1)), & &g_n=(-1)^n\exp(q_{n+1}(0)),\\
&p_n=u_{n,x}(-1)+u_{n+1,x}(0), & &\nu=\nu(x)=\exp\left(-\int(1+\mu(x))dx\right).
\end{align*}
\end{proposition}
\begin{remark}
\rm
It is easy to verify that in the semidiscrete case the change of variables $h_n(j)\to q_n(j)$
could be performed modulo addition of an arbitrary function to all dynamical variables $q_n(j)$.
Besides this, if $q_n(j)$ is a solution to the system~(\ref{todaq}), then $q_n(j)+\eps$ is also
a solution to this system for an arbitrary function $\eps=\eps(x)$. This means that having all
solutions of the half-infinite Toda lattice with the boundary condition
\begin{equation}
\label{modclos}
q_{n,x}(-1)+q_{n,x}(0)-h_n(0)=0,
\end{equation}
one can find all solutions of this lattice satisfying the constraint~(\ref{nint}) for an
arbitrary function $\mu$ by a proper choice of a function $\eps$. Therefore we'll assume that
$\mu=0$ for a $C$-series semidiscrete Toda lattice further on. This means that we'll consider
a bit more rigid constraint~(\ref{modclos}) instead of the constraint~(\ref{todac}). This ambiguity
in the change of variables $h_n(j)\to q_n(j)$ also leads to the necessity to introduce an awkward
multiplier $\nu$ in the entries of Lax matrices as well.
\end{remark}
\begin{remark}
\rm
Lax presentation for semidiscrete Toda lattice with boundary conditions
$$
q_{n,x}(-1)+q_{n,x}(0)-h_n(0)=0,\quad q_{n,x}(r+1)+q_{n,x}(r)-h_n(r)=0
$$
for a certain $r\geqslant 1$ is obtained similarly.
\end{remark}

\section{Symmetry approach in the semidiscrete case}\label{discsymm}

Symmetries of two-dimensional Toda lattice in the semidiscrete case as well as in the continuous case are
expressed in terms of nonlocal variables. We'll state the basic propositions and formulas for the semidiscrete
case that are similar to the ones from the Section~\ref{contsymm}. Introduce the following notation: $\pa_n=I-T$,
where $I$ is the identity operator and $b_n (j)=q_{n,x}(j)$. Hence the lattice~(\ref{todaq}) can be rewritten as
follows:
\begin{equation}
\label{aaa}
\pa_n b_n (j)=h_n(j)-h_n(j-1).
\end{equation}
Define the nonlocal variables $b_n^{(1)}(j)$ by the following formulas:
$$
\pa_x b_n (j)=b_n^{(1)}(j)-b_n^{(1)}(j-1),\quad\pa_n b_n^{(1)} (j)=\pa_x h_n(j).
$$
Compatibility of these equations is provided by the Toda lattice~(\ref{aaa}):
$$
\pa_n\pa_x b_n (j)=\pa_n \left(b_n^{(1)}(j)-b_n^{(1)}(j-1)\right)=\pa_x \left(h_n (j)-h_n (j-1)\right)=\pa_x\pa_n b_n (j).
$$
The next step is to determine the derivatives $\pa_x b_n^{(1)}(j)$ but this, however, requires to introduce nonlocalities $b_n^{(2)} (j)$
of the second order. Let
$$
\pa_x b_n^{(1)} (j)=b_n^{(1)}(j)\left(b_{n+1}(j+1)-b_n(j)\right)+h_n(j)\left(b_n^{(1)}(j-1)-b_n^{(1)}(j)\right)+b_n^{(2)}(j)-b_n^{(2)}(j-1);
$$
then the compatibility condition $\pa_n\pa_x b_n^{(1)}(j)=\pa_x\pa_n b_n^{(1)}(j)$ obviously leads to the following relation:
$$
\pa_n b_n^{(2)}(j)=h_n(j)b_{n+1}^{(1)}(j+1)-h_{n+1}(j+1)b_{n+1}^{(1)}(j).
$$
The following Proposition is proved by standard inductive reasoning.
\begin{proposition}
Nonlocal variables $b_n^{(k)}(j)$, where $k=2,3,\dots$, satisfy the following equations:
\begin{eqnarray}
\label{bbb}
\left\lbrace
\begin{array}{lll}
\pa_n b_n^{(k)}(j)&=&h_n (j)b_{n+1}^{(k-1)}(j+1)-h_{n+k-1} (j+k-1)b_{n+1}^{(k-1)}(j)\\
\pa_x b_n^{(k)}(j)&=&b_n^{(k)}(j)\left(b_{n+k}(j+k)-b_n(j)\right)+\\
&+&h_{n+k-1} (j+k-1)\left(b_n^{(k)}(j-1)-b_n^{(k)}(j)\right)+b_n^{(k+1)}(j)-b_n^{(k+1)}(j-1)
\end{array}
\right.
\end{eqnarray}
The compatibility of these equations for each $k$ is provided by the first of them, but for $k+1$.
\end{proposition}

In the semidiscrete case symmetries of the Toda lattice are defined as follows.
\begin{proposition}
The flows
\begin{eqnarray}
\label{dsymm2}
q_{n,t}(j)&=&b_n^2 (j)+b_n^{(1)}(j)+b_n^{(1)}(j-1),\\
\nonumber
q_{n,t}(j)&=&b_n^3 (j)+b_n^{(2)}(j)+b_n^{(2)}(j-1)+b_n^{(2)}(j-2)+\\
\nonumber
&+&b_n^{(1)}(j)\left(2b_n(j)+b_n(j+1)\right)+b_n^{(1)}(j-1)\left(2b_n(j)+b_n(j-1)\right)-\\
\label{dsymm3}
&-&b_n^{(1)}(j)h_n(j+1)-b_n^{(1)}(j-1)h_n(j)-b_n^{(1)}(j-2)h_n(j-1)
\end{eqnarray}
define symmetries of the lattice~(\ref{todaq}).
\end{proposition}

New set of dynamic variables that are more convenient for our approach is introduced in the same way as in the continuous case:
\begin{equation}
\label{qqq}
u_n=e^{-q_n(-2)},\quad v_n=e^{q_{n+1}(-1)},\quad w_n=q^{-q_n(0)},\quad z_n=e^{q_{n+1}(1)}.
\end{equation}
Replace the nonlocal variables $b_n^{(2)}(-2)$, $b_n^{(1)}(-2)$, $b_n^{(2)}(0)$ and $b_n^{(1)}(0)$
as follows:
\begin{align}
\notag
&r_n=b_n^{(1)}(-2),&&s_n=b_n^{(2)}(-2)+r_n (b_n(-1)-h_n(-1)),\\
\notag
&\rho_n=b_n^{(1)}(0),&&\sigma_n=b_n^{(2)}(0)+\rho_n (b_n(1)-h_n(1)).
\end{align}
The following Proposition is proved by straightforward calculation.
\begin{proposition}
In terms of the new variables the symmetry~(\ref{dsymm2}) is expressed as follows:
\begin{equation}
\label{symm2n}
u_{n,t}=u_{n,xx}+2r_n u_n,\quad v_{n,t}=-v_{n,xx}-2r_n v_n,\quad w_{n,t}=w_{n,xx}+2\rho_n w_n,\quad z_{n,t}=-z_{n,xx}-2\rho_n z_n,
\end{equation}
and the symmetry~(\ref{dsymm3}) can be rewritten in the following way:
\begin{align}
\label{symm3n}
&u_{n,t}=u_{n,xxx}+3r_n u_{n,x}-3s_n u_n+3r_{n,x}u_n,&& v_{n,t}=v_{n,xxx}+3r_{n+1}v_{n,x}+3s_{n+1}v_n\\
\label{symm3nn}
&w_{n,t}=w_{n,xxx}+3\rho_n w_{n,x}-3\sigma_n w_n+3\rho_{n,x}w_n,&& z_{n,t}=z_{n,xxx}+3\rho_{n+1}z_{n,x}+3\sigma_{n+1}z_n.
\end{align}
\end{proposition}

These formulas are more convenient then the formulas~(\ref{dsymm2},\ref{dsymm3}) because of their compactness. Besides this,
the flow defining the symmetry for each of the dynamic variables is expressed {\it only} in terms of {\it this} variable and
two nonlocalities. And, finally, the first of the identities~(\ref{bbb}) has a very compact form in terms of these variables.
The following Proposition holds.
\begin{proposition}
Difference derivatives of the functions $r_n$, $s_n$, $\rho_n$, $\sigma_n$ are total derivatives with respect to $x$:
\begin{align}
\label{nnn}
& \pa_n r_n=\pa_x (u_n v_n), &&\pa_n s_n=\pa_x\left(v_{n,x}u_n-\frac{1}{2}(u_n v_n)^2\right),\\
\label{ppp}
& \pa_n\rho_n=\pa_x (w_n z_n), &&\pa_n\sigma_n=\pa_x\left(z_{n,x}w_n-\frac{1}{2}(w_n z_n)^2\right).
\end{align}
\end{proposition}

We are going to find the cut-off constraints for the infinite semidiscrete Toda lattice that are compatible with the
symmetries~(\ref{dsymm2}) and~(\ref{dsymm3}). For convenience, this should be done in terms of the new set of variables~(\ref{qqq})
rather than in terms of the old one. In theory, it is possible to consider as complicate cut-off constraints as one wants;
but from the practical point of view of obtaining a complete description of all integrable cut-off constraints of a certain form,
we'll restrict ourselves by the following case. We'll examine the relations between four dynamic variables consecutive in the variable
$j$ and their derivatives with respect to $n$ and with respect to $x$. Without loss of generality one may assume that $j=-2,-1,0,1$,
i.e. that we are looking for a relation between the variables $u_n$, $v_n$, $w_n$, $z_n$ and their derivatives. However, cut-off
constraints may impose additional relations between nonlocal variables (as it happens in the continuous case) and these relations
may be as complicate as one wants as well. Therefore here we also have to restrict ourselves by some reasonable limitations on the
relations considered. Indeed, we'll assume that the nonlocal variables $r_n$, $s_n$, $\rho_n$ and $\sigma_n$ are independent from all
dynamic variables except for
$$
u_n,\ v_n,\ w_n,\ z_n,\ u_{n\pm 1},\ v_{n\pm 1},\ w_{n\pm 1},\ z_{n\pm 1},\dots\footnote{Nonlocal variables
$r_{n\pm 1}$, $\rho_{n\pm 1}$, $s_{n\pm 1}$, $\sigma_{n\pm 1}$,\dots are expressed in term of the dynamic variables and nonlocalities
$r_n$, $\rho_n$, $s_n$, $\sigma_n$ using the equations~(\ref{nnn},\ref{ppp}).}
$$
Under the above assumptions the following Propositions hold.
\begin{proposition}
\it
The only cut-off constraint of the form $u_n=F(v_{n-1},v_n,w_n,w_{n+1},z_n,z_{n+1})$ compatible with the symmetry~(\ref{symm2n}) is the
trivial one: $u_n=0$.
\end{proposition}
\begin{proposition}
\it
If the dynamic variables
\begin{equation}
\label{amm}
v_{n-1},\quad v_n,\quad w_n,\quad w_{n+1},\quad z_n,\quad z_{n+1},\quad v_{n-1,x},\quad w_{n,x},\quad z_{n,x}
\end{equation}
are independent, then among all cut-off constraints of the form $u_n=F(v_{n-1},v_n,w_n,w_{n+1},z_n,z_{n+1})$ the only constraint compatible with
the symmetry~(\ref{symm3n},\ref{symm3nn}) is the trivial one: $u_n=0$.
\end{proposition}

Therefore in order to find non-trivial boundary conditions compatible with the symmetry~(\ref{symm3n},\ref{symm3nn}) one has to impose a relation
between the variables~(\ref{amm}). The direct but rather cumbersome calculation leads to the following Theorem.
\begin{theorem}
\it
Let the dynamic variables~(\ref{amm}) be linked by a relation of the form
$$
v_{n-1,x}=H(v_{n-1},v_n,w_n,w_{n+1},z_n,z_{n+1},w_{n,x},z_{n,x})
$$
and suppose a non-trivial cut-off constraint of the form $u_n=F(v_{n-1},v_n,w_n,w_{n+1},z_n,z_{n+1})$ is compatible with the
symmetry~(\ref{symm3n},\ref{symm3nn}). Then
\begin{equation}
\label{sss}
H=\frac{v_{n-1}w_{n,x}}{w_n}+v_{n-1}z_n w_n,
\end{equation}
and cut-off constraint is as follows:
\begin{equation}
\label{rrr}
u_n=\frac{z_{n+1}w_{n+1}}{v_n}.
\end{equation}
\end{theorem}

Obviously, the cut-off constraint~(\ref{rrr}) is exactly the same as the boundary condition~(\ref{todac}) and the relation~(\ref{sss}) is equivalent
to the relation~(\ref{modclos}) that has already appeared before. This means that in the semidiscrete case as well as in the continuous one the natural
class of cut-off constraints
of the form $u_n=F(v_{n-1},v_n,w_n,w_{n+1},z_n,z_{n+1})$ leads to finite Toda lattices corresponding to classical simple Lie algebras (more precisely,
to their semidiscrete analogs).
\begin{remark}
\rm
The cut-off constraints on the right edge compatible with symmetries~(\ref{symm2n}) or~(\ref{symm3n},\ref{symm3nn}) are examined similarly.
\end{remark}

\section{Acknowledgements}

I am grateful to V.~Adler who drove my attention to the problem of classification of discrete analogs of two-dimensional
Toda lattices and who communicated me many facts that appeared to be useful for this work. Also I'd like to thank A.~Shabat for useful discussions.

The work is partially supported by the Russian President grant HSh-4995.2012.1, Russian Government grant 2010-220-01-077 and RFBR grants no.~11-01-00197-a
and 11-01-12067-ofi-m.

\end{document}